\documentstyle[twoside]{article}

\catcode`\@=11
\long\def\@makefntext#1{
\protect\noindent \hbox to 3.2pt {\hskip-.9pt  
$^{{\eightrm\@thefnmark}}$\hfil}#1\hfill}		%CAN BE USED 

\def\@makefnmark{\hbox to 0pt{$^{\@thefnmark}$\hss}}	%ORIGINAL 
	
\def\ps@myheadings{\let\@mkboth\@gobbletwo
\def\@oddhead{\hbox{}
\rightmark\hfil\eightrm\thepage}   
\def\@oddfoot{}\def\@evenhead{\eightrm\thepage\hfil
\leftmark\hbox{}}\def\@evenfoot{}
\def\sectionmark##1{}\def\subsectionmark##1{}}

\oddsidemargin=\evensidemargin
\newcounter{sectionc}\newcounter{subsectionc}\newcounter{subsubsectionc}
\renewcommand{\section}[1] {\vspace{12pt}\addtocounter{sectionc}{1} 
\setcounter{subsectionc}{0}\setcounter{subsubsectionc}{0}\noindent 
	{\tenbf\thesectionc. #1}\par\vspace{5pt}}
\renewcommand{\subsection}[1] {\vspace{12pt}\addtocounter{subsectionc}{1} 
	\setcounter{subsubsectionc}{0}\noindent 
	{\bf\thesectionc.\thesubsectionc. {\kern1pt \bfit #1}}\par\vspace{5pt}}
\renewcommand{\subsubsection}[1] {\vspace{12pt}\addtocounter{subsubsectionc}{1}
	\noindent{\tenrm\thesectionc.\thesubsectionc.\thesubsubsectionc.
	{\kern1pt \tenit #1}}\par\vspace{5pt}}
\newcommand{\nonumsection}[1] {\vspace{12pt}\noindent{\tenbf #1}
	\par\vspace{5pt}}

\newcounter{appendixc}
\newcounter{subappendixc}[appendixc]
\newcounter{subsubappendixc}[subappendixc]
\renewcommand{\thesubappendixc}{\Alph{appendixc}.\arabic{subappendixc}}
\renewcommand{\thesubsubappendixc}
	{\Alph{appendixc}.\arabic{subappendixc}.\arabic{subsubappendixc}}

\renewcommand{\appendix}[1] {\vspace{12pt}
        \refstepcounter{appendixc}
        \setcounter{figure}{0}
        \setcounter{table}{0}
        \setcounter{lemma}{0}
        \setcounter{theorem}{0}
        \setcounter{corollary}{0}
        \setcounter{definition}{0}
        \setcounter{equation}{0}
        \renewcommand{\thefigure}{\Alph{appendixc}.\arabic{figure}}
        \renewcommand{\thetable}{\Alph{appendixc}.\arabic{table}}
        \renewcommand{\theappendixc}{\Alph{appendixc}}
        \renewcommand{\thelemma}{\Alph{appendixc}.\arabic{lemma}}
        \renewcommand{\thetheorem}{\Alph{appendixc}.\arabic{theorem}}
        \renewcommand{\thedefinition}{\Alph{appendixc}.\arabic{definition}}
        \renewcommand{\thecorollary}{\Alph{appendixc}.\arabic{corollary}}
        \renewcommand{\theequation}{\Alph{appendixc}.\arabic{equation}}
        \noindent{\tenbf Appendix \theappendixc #1}\par\vspace{5pt}}
\newcommand{\subappendix}[1] {\vspace{12pt}
        \refstepcounter{subappendixc}
        \noindent{\bf Appendix \thesubappendixc. {\kern1pt \bfit #1}}
	\par\vspace{5pt}}
\newcommand{\subsubappendix}[1] {\vspace{12pt}
        \refstepcounter{subsubappendixc}
        \noindent{\rm Appendix \thesubsubappendixc. {\kern1pt \tenit #1}}
	\par\vspace{5pt}}

\newcommand{\textlineskip}{\baselineskip=13pt}
\newcommand{\smalllineskip}{\baselineskip=10pt}

\def\eightcirc{
\begin{picture}(0,0)
\put(4.4,1.8){\circle{6.5}}
\end{picture}}
\def\eightcopyright{\eightcirc\kern2.7pt\hbox{\eightrm c}} 

\newcommand{\copyrightheading}[1]
	{\vspace*{-2.5cm}\smalllineskip{\flushleft
	{\footnotesize International Journal of Modern Physics A, #1}\\
	{\footnotesize $\eightcopyright$\, World Scientific Publishing
	 Company}\\
	 }}

\def\abstracts#1#2#3{{
	\centering{\begin{minipage}{4.5in}\baselineskip=10pt\footnotesize
	\parindent=0pt #1\par 
	\parindent=15pt #2\par
	\parindent=15pt #3
	\end{minipage}}\par}} 

\renewenvironment{thebibliography}[1]
	{\frenchspacing
	 \ninerm\baselineskip=11pt
	 \begin{list}{\arabic{enumi}.}
	{\usecounter{enumi}\setlength{\parsep}{0pt}
	 \setlength{\leftmargin 12.7pt}{\rightmargin 0pt} %FOR 1--9 ITEMS
	 \setlength{\itemsep}{0pt} \settowidth
	{\labelwidth}{#1.}\sloppy}}{\end{list}}

\newcounter{itemlistc}
\newcounter{romanlistc}
\newcounter{alphlistc}
\newcounter{arabiclistc}

\newcommand{\fcaption}[1]{
        \refstepcounter{figure}
        \setbox\@tempboxa = \hbox{\footnotesize Fig.~\thefigure. #1}
        \ifdim \wd\@tempboxa > 5in
           {\begin{center}
        \parbox{5in}{\footnotesize\smalllineskip Fig.~\thefigure. #1}
            \end{center}}
        \else
             {\begin{center}
             {\footnotesize Fig.~\thefigure. #1}
              \end{center}}
        \fi}

\newcommand{\tcaption}[1]{
        \refstepcounter{table}
        \setbox\@tempboxa = \hbox{\footnotesize Table~\thetable. #1}
        \ifdim \wd\@tempboxa > 5in
           {\begin{center}
        \parbox{5in}{\footnotesize\smalllineskip Table~\thetable. #1}
            \end{center}}
        \else
             {\begin{center}
             {\footnotesize Table~\thetable. #1}
              \end{center}}
        \fi}

\def\@citex[#1]#2{\if@filesw\immediate\write\@auxout
	{\string\citation{#2}}\fi
\def\@citea{}\@cite{\@for\@citeb:=#2\do
	{\@citea\def\@citea{,}\@ifundefined
	{b@\@citeb}{{\bf ?}\@warning
	{Citation `\@citeb' on page \thepage \space undefined}}
	{\csname b@\@citeb\endcsname}}}{#1}}

\newif\if@cghi
\def\cite{\@cghitrue\@ifnextchar [{\@tempswatrue
	\@citex}{\@tempswafalse\@citex[]}}
\def\citelow{\@cghifalse\@ifnextchar [{\@tempswatrue
	\@citex}{\@tempswafalse\@citex[]}}
\def\@cite#1#2{{$\null^{#1}$\if@tempswa\typeout
	{IJCGA warning: optional citation argument 
	ignored: `#2'} \fi}}

\def\pmb#1{\setbox0=\hbox{#1}
	\kern-.025em\copy0\kern-\wd0
	\kern.05em\copy0\kern-\wd0
	\kern-.025em\raise.0433em\box0}

\def\fnt#1#2{\footnotetext{\kern-.3em
	{$^{\mbox{\scriptsize #1}}$}{#2}}}

\def\fpage#1{\begingroup
\voffset=.3in
\thispagestyle{empty}\begin{table}[b]\centerline{\footnotesize #1}
	\end{table}\endgroup}

\def\runninghead#1#2{\pagestyle{myheadings}
\markboth{{\protect\footnotesize\it{\quad #1}}\hfill}
{\hfill{\protect\footnotesize\it{#2\quad}}}}
\font\tenrm=cmr10
\font\tenit=cmti10 
\font\tenbf=cmbx10
\font\bfit=cmbxti10 at 10pt
\font\ninerm=cmr9

\font\eightrm=cmr8

\def\qed{\hbox{${\vcenter{\vbox{			%HOLLOW SQUARE
   \hrule height 0.4pt\hbox{\vrule width 0.4pt height 6pt
   \kern5pt\vrule width 0.4pt}\hrule height 0.4pt}}}$}}

\input epsf.tex
\def\DESepsf(#1 width #2){\epsfxsize=#2 \epsfbox{#1}}
\begin{document}

\runninghead{Yukawa Textures in $\ldots$}
{Yukawa Textures in $\ldots$}

\normalsize\textlineskip
\thispagestyle{empty}
\setcounter{page}{1}

\copyrightheading{}			%{Vol. 0, No. 0 (1993) 000--000}

\vspace*{0.88truein}

\fpage{1}
\centerline{\bf Yukawa Textures in }
\vspace*{0.035truein}
\centerline{\bf Heterotic M-Theory}
\vspace*{0.37truein}
\centerline{\footnotesize R. Arnowitt and B. Dutta}
\vspace*{0.015truein}
\centerline{\footnotesize\it Department of Physics, Texas A\&M University}
\baselineskip=10pt
\centerline{\footnotesize\it College Station, TX 77843-4242,
USA}
%\vspace*{0.225truein}
%\publisher{(received date)}{(revised date)}

\vspace*{0.21truein}
\abstracts{We examine the structure of the Yukawa
couplings in the 11 dimensional Horava-Witten M-theory based on non-standard
embeddings. We find that  the
CKM and quark mass hierarchies can be explained in M Theory without introducing undue 
fine tuning.  A
phenomenological example is presented satisfying all CKM and quark mass data
requiring  the 5-branes cluster near the second orbifold plane, and that the
instanton charges of the physical orbifold plane vanish. The latter condition is
explicitly realized on a Calabi-Yau manifold with del Pezzo base $dP_7$.}{}{}

%\textlineskip			%) USE THIS MEASUREMENT WHEN THERE IS
%\vspace*{12pt}			%) NO SECTION HEADING
\textlineskip\vspace*{12pt}	 In Horava-Witten heterotic M-theory \cite{hw1} with
``non-standard" embeddings  space has an 11 dimensional orbifold structure of the form
(to lowest order) $M_4\times X\times S^1/Z_2$ where $M_4$ is Minkowski space, $X$ is a 6
dimensional (6D) Calabi-Yau space, and $-\pi\rho\leq x^{11}\leq\pi \rho$. There are two
orbifold 10D manifolds $M_4\times X$ at the $Z_2$ fixed points at $x^{11}=0$ and
$x^{11}=\pi\rho$, each with an a priori $E_8$ gauge symmetry. Physical matter lives on
the $x^{11}=0$ orbifold plane and only gravity lives in the bulk.

 In addition there can be a set of 5-branes in the bulk at points
$0< x_n<\pi\rho$, $n=1...N$ each spanning $M_4$ (to preserve Lorentz invariance) and
wrapped on a holomorphic curve in $X$ (to preserve N=1 supresymmetry).   The existence of
these 5-branes allows one to satisfy the cohomological constraints with $E_8$ on the
$x^{11}=0$ plane breaking to $G\times H$ where $G$ is the structure group of the
Calabi-Yau manifold and $H$ is the physical grand unification group.  We consider here
the case
$G=SU(5)$, and hence $H=SU(5)$.

Recently, three generation models with a Wilson line
breaking $SU(5)$ to $SU(3)\times SU(2)\times U(1)$ have been constructed in the M-theory
frame work using torus fibered Calabi-Yau manifolds (with two sections)\cite{dopw}. Also,
the general structure of the Kahler metric of the matter field has
been examined\cite{low4}. We find that this structure can lead to
Yukawa textures with all CKM and quark mass data in agreement with experiment.
  We have also
shown that  a three generation model with a Wilson line (to break SU(5) to the standard
model) and which  explains the hierarchies  in the matrix 
elements of the Yukawa sector can be constructed
on a Calabi-Yau manifold with del-Pezzo base
$dP_7$ \cite{ad}.

 The bose part of the 11 dimensional gravity multiplet consists of
the metric tensor $g_{IJ}$, the antisymmetric 3-form $C_{IJK}$ and its field strength
$G_{IJKL}=24
\partial_{[I}C_{JKL]}$. ($I,\,J,\,J,\,K=1...11.$). The $G_{IJKL}$ obey field equations
$D_IG^{IJKL}=0$ and Bianchi identities
\begin{eqnarray}\nonumber(dG)_{11RSTU}&=&4\sqrt 2\pi({\kappa\over
{4\pi}})^{2/3}[J^0\delta(x^{11})\\\nonumber+J^{N+1}\delta(x^{11}-\pi \rho) &+&{1\over
2}\Sigma^N_{n=1}J^{n}(\delta(x^{11}-x_n)\\&+&\delta(x^{11}+ x_n))]_{RSTU}
\label{eq3}\end{eqnarray} Here $(\kappa^{2/9})$ is the 11 dimensional Planck scale, and
$J^{n}$,
$n=0,\,1,\,...N+1$ are sources from orbifold planes and the $N$ 5-branes. These equations
can be solved perturbatively in powers of $(\kappa^{2/3})$
\cite{low4}. The effective 4D theory is then given by a Kahler potential
$K=Z_{IJ}{\bar C^I}C^J$, Yukawa couplings $Y_{IJK}$ for the matter fields $C^{I}$ and
gauge functions from the physical orbifold plane $x^{11}$=0. To first order, $Z_{IJ}$
takes the form \cite{low4}. 
\begin{equation}  Z_{IJ}=e^{-K_T/3}[G_{IJ}-{\epsilon\over{2V}}\tilde{\Gamma}^i_{IJ}
\Sigma^{N+1}_0(1-z_n)^2\beta^{(n)}_i]
\label{eq2}\end{equation}$\epsilon=(\kappa/4\pi)^{2/3}2\pi^2\rho/V^{2/3}$ is the
expansion parameter. $V$ is the Calabi-Yau volume, $G_{IJ}$, $\tilde\Gamma^i_{IJ}$ and
$Y_{IJK}$ can be expressed in terms of integrals over the Calabi-Yau manifold, and $K_T$ is
the Kahler potential for the moduli.

The second term of Eq.(2) is  a small correction in order to make the perturbation analysis
to work. A priori one expects $G_{IJ}$, $\tilde\Gamma^{i}_{IJ}$ and
$Y_{IJK}$ to be of $O(1)$, and the parameter $\epsilon$ is not too
small. However, the second term will be small if $\beta^{(0)}_i$  vanishes and if
the 5-branes cluster near the distant orbifold plane i.e.
$d_n\equiv1-z_n$ is small, where $z_n=x_n/{\pi\rho}$. In the following we will assume
then that
\begin{equation}\beta^{(0)}_i=0;\,\,d_n=1-z_n\cong0.1\end{equation} The condition
$\beta^{(0)}_i=0$ is quite non trivial, but it is possible to show that a three generation
model of a torus fibered Calabi-Yau manifold with Wilson like breaking $SU(5)$ to
$SU(3)\times SU(2)\times U(1)$ with del -Pezzo base $dP_7$ has this property\cite{ad}.

Since $d_n$ is small, we will assume assume that the  $\epsilon$ term of Eq.(2) constitute the third generation
contributions to the Kahler metric.  A simple phenomenological
example for the u and d quark contributions with these properties (and containg the maximum
numbers of zeros) is $(f_T\equiv exp(-K_T/3))$:
\begin{eqnarray}\nonumber Z^u&=&f_T\left(\matrix{ 1  & 0.345  & 0 \cr 0.345 & 0.132 &
0.639 d^2
\cr 0  & 0.639 d^2& 0.333 d^2 }\right); \\ Z^d&=&  f_T\left(\matrix{ 1 &  0.821  & 0 \cr
0.821 & 0.887 & 0 \cr 0  & 0& 0.276  }\right).
\label{eq27}\end{eqnarray} with Yukawa matrices ${\rm diag}Y^u$={(0.0765, 0.536, 0.585
$Exp[
\pi i/2]$)} and ${\rm diag}Y^d$={(0.849,\, 0.11,\, 1.3)}.

To obtain the physical Yukawa matrices, one must first diagonalize the Kahler metric and
then rescale it to unity. Then using the renormalization group equations, one can
generate the CKM matrix, and the quark masses. The results are given in the following
table:\vspace{0.7cm}
\begin{center} \begin{tabular}{|c|c|c|}  
 \hline Quantity&Th. Value&Exp. Value\cite{com}\\\hline
$m_t$(pole)&170.5& 175$\pm$ 5\\
$m_c$($m_c$)&1.36&1.1-1.4\cr
$m_u$(1 GeV)& 0.0032&0.002-0.008\\
$m_b$($m_b$)& 4.13&4.1-4.5\cr
$m_s$(1 GeV)& 0.110&0.093-0.125\cite{sa}\\
$m_d$(1 GeV)& 0.0055&0.005-0.015\\
$V_{us}$&0.22&0.217-0.224\\
$V_{cb}$&0.036&0.0381$\pm$0.0021\cite{ss}\\
$V_{ub}$&0.0018&0.0018-0.0045\\
$V_{td}$&0.006&0.004-0.013\\
\hline\end{tabular}\end{center}\vspace{0.6cm} and sin$2\beta$=0.31 and sin$\gamma$=0.97.
The agreement with experiment is quite good. The quark mass ratios for the first two
generations are given as  
$m_u/m_d$=0.582 and
$m_s/m_d$=20.0. These values are in good agreement with Leutwyler evaluations\cite{leut}
0.553$\pm$ 0.043 and 18.9$\pm$0.8.

 Though the entries in $Z^{u,d}$ and $Y^{u,d}$ are chosen to obtain a precise fit to
the experimental results, as shown in
\cite{ad}, the quark mass hierarchies arise naturally from a Kahler metric of the type of Eq.
(4). Similarly the smallness of the off diagonal
$V_{CKM}$ matrix elements also occur
 naturally as a consequence of the above model. Thus it is possible for  M-theory to
generate the Yukawa hierarchies without any undue fine tuning and without introducing ad
hoc very small off diagonal entries.

M theory with non-standard embeddings is a new possibility of encoding
the Yukawa hierarchies in the Kahler metric. This can happen naturally if the
5-branes cluster near the hidden orbifold plane ($d_n\equiv1-z_n\simeq0.1$) and
the instanton charges of the physical plane vanish ($\beta^{(0)}_i=0$). 

This work was supported in part by NSF grant no. PHY-9722090.  

\nonumsection{References}


\begin{thebibliography}{99}
\bibitem{hw1} P. Horava and E. Witten, Nucl. Phys. B {\bf 460}, 506 (1996); Nucl. Phys. B
{\bf 475}, 94 (1996); E. Witten, Nucl. Phys. B {\bf 471}, 135 (1996).

\bibitem{dopw} R. Donagi, B. Ovrut, T. Pantev and D. Waldram, hep-th/9912208.

\bibitem{low4}A. Lukas, B. Ovrut and D. Waldram, JHEP {\bf 9904}, 009 (1999).

\bibitem{ad}R. Arnowitt and B. Dutta, hep-ph/0006172 (to appear in Nucl. Phys. B).

\bibitem{com} Except as otherwise noted, expermental entries are from the Particle Data
group, Eupropean  Phys. Journ. C {\bf 3}, 1 (1998).

\bibitem{sa}S. Aoki, hep-ph/9912288.

\bibitem{ss}S. Stone, hep-ph/9904350.
\bibitem{leut}H. Leutwyler, Phys. Lett. B {\bf 374}, 163 (1996).
\end{thebibliography}
\end{document}